\documentclass[superscriptaddress,twocolumn,showpacs,prl,amsmath,amssymb]{revtex4}

\usepackage[varg]{txfonts}

\usepackage{graphicx}

\begin{document}

\title{Expansion of a mesoscopic Fermi system from a harmonic trap}

\author{Pavel Nagornykh} \affiliation{Joint Quantum Institute and Condensed Matter Theory Center, Department
    of Physics, University of Maryland, College Park, MD 20742-4111}
\author{Victor Galitski} \affiliation{Joint Quantum Institute and Condensed Matter Theory Center, Department
    of Physics, University of Maryland, College Park, MD 20742-4111}
\begin{abstract}
We study quantum dynamics of an atomic Fermi system with a finite
number of particles, $N$, after it is released from a harmonic
trapping potential. We consider two different initial states: The
Fermi sea state and the paired state described by the projection
of the grand-canonical BCS wave function to the subspace with a
fixed number of particles. In the former case, we derive exact and
simple analytic expressions for the dynamics of particle density
and density-density correlation functions, taking into account the
level quantization and possible anisotropy of the trap. In the
latter case of a paired state, we obtain analytic expressions for
the density and its correlators in the leading order with respect
to the ratio of the trap frequency and the superconducting gap
(the ratio assumed small). We discuss several dynamic features,
such as time evolution of the peak due to pair correlations, which
may be used to distinguish between the Fermi sea and the paired
state.
\end{abstract}

 \maketitle \vspace*{-0.17in}

{\em Introduction ---} Measurements of density distributions and
noise correlations in time-of-flight expansion  have proven to be
a very powerful  direct probe of the quantum states of cold atomic
systems. In bosonic systems, density distributions contain
valuable information about the initial quantum state and allow one
to observe directly  Bose-Einstein condensates. In cold Fermi
systems, the time-of-flight density profile of a Fermi liquid is
expected to be qualitatively similar to the one of a paired state.
Altman et al.~\cite{Altman_etal} have shown that to identify the
existence of a BCS condensate, one should probe noise
correlations, which in the condensed state would acquire a peak at
the opposite momenta corresponding to Cooper pairing. Greiner et
al.~\cite{Greiner_etal} have experimentally observed similar
pairing correlations of fermionic atoms  on the BEC side of a
Feshbach resonance, showing the experimental feasibility of the
proposed method. However, so far there have been no direct
observations of pairing correlations in the BCS state, which was
detected in Ref.~[\onlinecite{schunck-2006}] by different means.

Most previous theoretical studies have concentrated on the dynamic
properties of density and its correlators, which followed from the
grand canonical wave-functions. However, all real experimentally
studied systems are finite and non-uniform due to the trapping
potential. Moreover, the grand canonical description is strictly
speaking not appropriate in this context, since non-trivial
density fluctuations may exist only if the system is finite and
formally vanish in thermodynamic limit. It is therefore important
to consider expansion dynamics of a finite system in a realistic
trapping potential. We should mention here that there exist a
number of exact results for a Fermi system in a harmonic
potential: See e.g., Ref.~[\onlinecite{Schneider}], which
considered thermodynamic properties of a mesoscopic Fermi gas and
Refs.~[\onlinecite{Brack,Wang1,Gleisberg,vanZyl,Bruun2}] where
density and energy distributions were studied. Also, Bruun and
Heiselberg~\cite{Bruun3} investigated the properties of a paired
state of trapped fermions.

In this work we extend these studies to the dynamic regime of an
expanding Fermi cloud. We explicitly calculate density
distributions and noise correlations using the canonical
formulation and taking into account discrete energy levels and a
finite number of particles in a harmonic trap. In the case of a
Fermi sea state, we derive exact and simple analytic expressions
for the density and its correlators. To describe a paired state,
we use the grand canonical BCS wave-function projected onto the
subspace with a fixed number of particles - projected BCS (PBCS)
state. This wave-function was written explicitly in the original
paper of Bardeen, Cooper, and Shrieffer~\cite{The_BCS} and then
used in the context of nanoscale superconductors~\cite{von_Delft}.
We should mention here that there exists an exact solution of the
reduced BCS Hamiltonian due to
Richardson~\cite{Richardson1,Richardson_Sherman}, who found a set
of equations, which determine the energy levels in a finite Fermi
system with attractive interactions and constructed an exact
wave-function in this case. However, to use this exact solution is
difficult due to the complicated structure of the Richardson's
equations. Moreover, it has been shown~\cite{von_Delft} that the
exact Richardson's wave function reduces to the PBCS state if the
superconducting gap $\Delta$ is much larger than the level spacing
(in fact the BCS Ansatz remains a good approximation even if the
ratio of the gap and level spacing is of order
unity~\cite{Braun,von_Delft}). Below, we consider the latter
limit, which in the context of a trapped Fermi gas corresponds to
$\Delta \gg \omega$, where $\omega$ is the frequency of the
trapping potential.

{\em Time-dependent density ---} Let us consider  a Fermi system
in a harmonic trap in a quantum many-body state $| \Phi \rangle$.
We are interested in the dynamics of density and its correlators
after the harmonic potential and interactions, if present, are
turned off at $t=0$. For simplicity, we start with the
one-dimensional case. We will see that generalization of all
results to higher dimensions is straightforward. The
time-dependent density is
\begin{equation}
\label{n1} \langle
n(x,t)\rangle=\langle\Phi|\hat{U}_0^{+}(t)\hat{\psi}^{+}(x)\hat{\psi}(x)\hat{U}_0(t)|\Phi\rangle,
\end{equation}
where $\hat{U}_0(t)=\exp\left(-i\sum_{p}E_p \hat{a}^{+}_p
\hat{a}_p\, t\right)$ is the evolution operator of the free
system, $E_p = p^2/(2m)$, and $\hat{\psi}(x)=\sum_n
\psi_n(x)\hat{a}_n$  is the field operator, where $\psi_n(x)$ and
$\hat{a}_n$ are the oscillator wave-function and the annihilation
operator in the harmonic potential corresponding to the $n$-th
level. Introducing the Fourier transform of the oscillator
wave-functions $\tilde{\psi}_n(p)$, we can write the density in
the following form
\begin{equation}
\label{n2} \langle n(x,t)\rangle=\sum_{n,m}\int
\frac{dq}{2\pi}\,\frac{dP}{2\pi}\, e^{iq \left( x - i\frac{P}{m}t
\right)}
\tilde{\psi}^{*}_n\left(P+\frac{q}{2}\right)\tilde{\psi}_m\left(P-\frac{q}{2}\right)
\langle \hat{a}^{+}_n \hat{a}_m\rangle.
\end{equation}
We note that if we neglect the $q$-dependence of the wave
functions (which is a good approximation if $t \gg \omega^{-1}$),
we recover the ``natural'' result for the
density~\cite{Altman_etal}
$$\langle n\rangle \approx
\sum_{n,m}\tilde\psi^{*}_n(P)\tilde\psi_m(P) \langle a^{+}_n
a_m\rangle \Bigr|_{P=mx/t}. $$
 We also note that Eq.~(\ref{n2}) is
exact and valid for any many-body state $|\Phi\rangle$, assuming
that interactions are absent at $t > 0$.

{\em Fermi gas ---} Now let us consider a Fermi sea state at zero
temperature, which means that $| \Phi\rangle =
\prod\limits_{n=0}^{N-1} \hat{a}^\dagger_n |{\rm vac}\rangle$. We
find
\begin{multline}
\langle n(x,t) \rangle=\sum\limits_{n=0}^{N-1}
\frac{2\pi}{2^nn!\sqrt{\pi}m\omega }\int
\,\frac{dp}{2\pi}\,\frac{dp'}{2\pi}\,H_n\left(\frac{p}{\sqrt{m\omega}}\right)
H_n\left(\frac{p'}{\sqrt{m\omega}}\right)\\
 \times e^{i(p-p')x} e^{i(E_{p'}-E_p)(t + \frac{i}{\omega})}.
\label{nFL1}
\end{multline}
The integrals over $p$ and $p'$ are table integrals and we obtain
\begin{eqnarray}
\label{nFL3}  \langle n(x,t) \rangle =\sum\limits_{n=0}^{N-1}
\psi_n^2\left[x,\omega(t)\right],
 \end{eqnarray}
where $\psi_n\left[x,\omega(t)\right]$ is the oscillator wave
function with a ``rescaled'' frequency $\omega(t) =
\omega/(1+\omega^2 t^2).$ The sum in Eq.~(\ref{nFL3}) can be
calculated exactly and we find
\begin{equation}
\label{nFL} \langle
n(x,t)\rangle=\frac{e^{-x^2/\xi^2}}{2^{N}(N-1)!\sqrt{\pi}\xi}
\left[H^2_{N}\left(\frac{x}{\xi}\right)-H_{N+1}\left(\frac{x}{\xi}\right)
H_{N-1}\left(\frac{x}{\xi}\right)\right],
\end{equation}
where we introduced $\xi =
{\sqrt{1+\omega^2t^2}}/{\sqrt{m\omega}}$. We see that the behavior
of the density is determined by the wave-functions near the
``Fermi level'' $n_{\rm F}=N-1$. One can also check by explicit
calculation that the integral over $x$ of Eq.~(\ref{nFL})
naturally reproduces the total number of particles in the system.
It is also easy to generalize the result (\ref{nFL3}) to the
three-dimensional case. In the case of an anisotropic trap with
frequencies $\omega_{x,y,z}$ in the corresponding directions, we
find
\begin{eqnarray}
\label{nFL3D}  \langle n({\bf r},t) \rangle
=\sum\limits_{E_{n_x,n_y,n_z} < E_{\rm F}}
\psi_{n_x}^2\left[x,\omega_x(t)\right]\psi_{n_y}^2\left[y,\omega_y(t)\right]
\psi_{n_z}^2\left[z,\omega_z(t)\right],
 \end{eqnarray}
where $E_{\rm F}$ is the Fermi level, $E_{n_x,n_y,n_z} = \omega
\left( n_x + n_y +n_z +3/2 \right)$, and $\omega_\alpha (t)=
\omega_\alpha/(1+\omega_\alpha^2 t^2)$. This result (\ref{nFL3D})
allows us to determine the dynamic  ratio of the radii of the
time-dependent density distribution. E.g., let us assume that
$\omega_x = \omega_y = \omega_\perp > \omega_z$. The dynamic
aspect ratio is determined by the ratio of the spreads of the
wave-functions corresponding to the fermions occupying the highest
possible levels $n_{x,z}^{\rm max} = E_{\rm F}/\omega_{x,z}$.
\begin{equation}
\label{ratio} \frac{R_\perp(t)}{R_z(t)} \sim \frac{\sqrt{n_x^{\rm
max}} \xi_\perp(t)}{\sqrt{n_z^{\rm max}} \xi_z(t)} =
\frac{\omega_z}{\omega_\perp}\, \frac{\sqrt{1 + \omega_\perp^2
t^2}}{\sqrt{1 + \omega_z^2 t^2}}.
\end{equation}
We therefore reproduce the classical formula of
Ref.~[\onlinecite{Menotti}]. According to Eq.~(\ref{ratio}),  the
ratio crosses over from the initial value
$\omega_z/\omega_\perp<1$ at $t=0$ to a final constant value 1 at
$t \to \infty$. We mention here that the dynamics of the single
particle wave-functions corresponding to low-lying levels of the
harmonic potential are different: The ratio of the spreads of the
corresponding wave-packets behaves as $\frac{\xi_\perp}{\xi_z} =
\frac{\sqrt{\omega_z}}{\sqrt{\omega_\perp}}\, \frac{\sqrt{1 +
\omega_\perp^2 t^2}}{\sqrt{1 + \omega_z^2 t^2}}$ and crosses over
from the initial value $\sqrt{\omega_z/\omega_\perp}<1$ at $t=0$
to a final constant value $\sqrt{\omega_\perp/\omega_z} > 1$ at $t
\to \infty$.

In the case of an isotropic three-dimensional trap, we can rewrite
Eq.~(\ref{nFL3D}) in the form
\begin{equation}
\label{nFLisotr}
 \langle n({\bf r},t) \rangle
=\sum\limits_{n<n_{\rm F},l}\,\,\,\, (2l+1)
R_{nl}^2\left[r,\omega(t)\right],
\end{equation}
where $R_{nl}(r,\omega)$ is the radial wave-function of a harmonic
potential with a frequency $\omega$,  $R_{nl}=C_{nl}
e^{-r^2/\xi^2}(r/\xi)^l F\left[-(n-l)/2,l+3/2,-r^2/\xi^2\right]$
with $F(\cdot)$ being the hypergeometric function.

We also mention here that generalization of the obtained results
to the case of a finite temperature is straightforward. Indeed,
Eq.~(\ref{nFLisotr}) describes a superposition of single-particle
wave-functions corresponding to the occupied states. At zero
temperature all states below the Fermi level are occupied with the
equal probability. At finite temperatures, we have to weight each
state with the probability determined by the Fermi distribution
function $f(E,T) = \left\{ 1 + \exp\left[\frac{E - \mu(T)}{T}
\right]\right\}^{-1}$, where $\mu(T)$ can be found from the
relation $\sum_n g_n f(E_n) = N$ with $g_n = (n+1)(n+2)/2$ being
the degeneracy of the $n$-th level. Thus, it is natural to write
the finite temperature result as
\begin{equation}
\label{nFLisotrT}
 \langle n({\bf r},t;T) \rangle
=\sum\limits_{n<n_{\rm F},l}\,\,\,\, (2l+1)
R_{nl}^2\left[r,\omega(t)\right] f(E_n,T).
\end{equation}

Similarly, one can obtain the density-density noise correlations
of a mesoscopic Fermi gas. Below, we just give a finite result of
the calculation in the one-dimensional case at $T=0$
\begin{eqnarray}
\nonumber &&\langle n(x_1,t)n(x_2,t)\rangle=\delta(x_1-x_2)\langle
n(x_1,t)\rangle+\langle n(x_1,t)\rangle\langle n(x_2,t)\rangle
\\&&
\label{nnFL}  - \frac{e^{
-\frac{x_1^2+x_2^2}{\xi^2}}\left[H_{N-1}\left(\frac{x_1}{\xi}\right)H_{N}\left(\frac{x_2}{\xi}\right)
-H_{N}\left(\frac{x_1}{\xi}\right)H_{N-1}\left(\frac{x_2}{\xi}\right)\right]^2}{\pi
(N-1)!^2\,2^{2N}(x_1-x_2)^2},
\end{eqnarray}
where $\langle n(x_{1,2},t) \rangle$ is given by Eq.~(\ref{nFL}).
We note that in thermodynamic limit the density correlations
vanish. We also mention that higher-order density correlation
functions of a Fermi gas can be obtained analytically and have a
structure similar to Eq.~(\ref{nnFL}).

{\em Paired state --- } Now we consider density dynamics of an
expanding finite Fermi system initially in the paired state. If
the system were infinite, we could use the  grand canonical BCS
wave-function. However, the latter is not appropriate in the given
context since the grand canonical BCS Ansatz  does not conserve
the number of particles.
 A superconducting wave-function with a
fixed number of particles was discussed by Bardeen, Cooper, and
Schrieffer in their seminal paper~\cite{The_BCS} and later in
great detail by von Delft~\cite{von_Delft} and others in the
context of ultrasmall superconducting grains. This issue was also
investigated by Richardson~\cite{Richardson1,Richardson_Sherman}
who found an exact many body wave function of the reduced
Hamiltonian
\begin{equation}
\label{RH}
 \hat{H} = \sum_{nn'} \left( \delta_{nn'}  \varepsilon_n \hat{a}_{n
\sigma}^\dagger  \hat{a}_{n' \sigma} - \lambda \omega \hat{a}_{n
\uparrow}^\dagger \hat{a}_{{n_T} \downarrow}^\dagger  \hat{a}_{n'
\downarrow} \hat{a}_{{n'_T} \uparrow} \right),
\end{equation}
where $|n\rangle$ and $|n_T\rangle = \hat{T} |n\rangle$ are pairs
of single particle states
 related to each other by time-reversal, the sum runs over all
interacting single-particle levels, and $\varepsilon_n$ are the
energies of the levels.  Model Hamiltonian (\ref{RH}) has been
used to describe the properties of nuclei in the nuclear pairing
model as well as small superconducting grains with a finite number
of particles. The reduced Richardson Hamiltonian should also be an
appropriate model to describe a mesoscopic paired Fermi system in
a trap, with $n$ corresponding to the quantum levels of the
harmonic trap. Note that in the one-dimensional case, the paired
states related to each other by time-reversal are identical
$|n\rangle=|n_T\rangle$, while in a three-dimensional potential
they are $|n,l,m\rangle$ and $ |n,l,-m\rangle = \hat{T}
|n,l,m\rangle$. The exact wave-function of reduced
Hamiltonian~(\ref{RH}) is
$$
|{\rm RBCS}\rangle_{\rm exact} = C_N' \prod\limits_\nu^{N} \sum_n
\frac{\hat{b}_n^\dagger}{2 \varepsilon_n -
E_\nu}\,|\mathrm{vac}\rangle,
$$
where $\hat{b}_n^\dagger = \hat{a}_{n \uparrow}^\dagger
\hat{a}_{{n_T} \downarrow}^\dagger $ and $E_\nu$ are solutions of
a system of coupled Richardson's
equations~\cite{Richardson1,Richardson_Sherman,von_Delft}. The
latter can not be solved analytically, but one can show that in
the limit of large but finite number of particles and small level
spacing the exact wave-function is  indistinguishable from the
variational BCS wave-function projected to the  fixed number of
particles. This projected BCS wave-function has the form
\begin{equation}
\label{PBCS} |{\rm PBCS}\rangle = C_N \int \frac{d \phi} {2 \pi}
e^{-i\phi N} \prod_n \left( u_n + e^{i\phi} v_n \hat{b}_n^\dagger
\right) |\mathrm{vac}\rangle,
\end{equation}
where $C_N$ is a normalization constant and $v_n$ and $u_n$ are
variational parameters. We now use this PBCS
wave-function~(\ref{PBCS}) to calculate the dynamics of the
density, which follows after the trap and the BCS interaction are
turned off. For simplicity, we study the one-dimensional case
[generalization to higher dimensions is formally straightforward;
see also  Eqs.~(\ref{nFL3D},~\ref{nFLisotr}) and the corresponding
discussion]. From Eqs.~(\ref{n2}) and (\ref{PBCS}), we obtain
after some algebra
\begin{equation}
\label{nBCS1} \langle n(x,t)\rangle=
\frac{2\sum\limits_{n=0}^\infty f_n |\psi_n[x,\omega(t)]|^2
\sum\limits_{\{m\}_{N-1}^n}\prod\limits_{m \in
{{\{m\}_{N-1}^n}}}f_m } {\sum\limits_{\{n\}_{N}}\prod\limits_{n
\in{\{n\}_{N}}}f_n},
\end{equation}
where  $\{n\}_{N}$ implies all possible sets of $N$ quantum
levels, $\{m\}_{N-1}^n$ denotes all possible sets of $N-1$ quantum
levels except $n$, $f_n = v_n^2/u_n^2$, and $\psi_n[x,\omega(t)]$
is the oscillator wave-function with a rescaled frequency
$\omega(t) = \omega/(1+\omega^2 t^2)$. In the limit $\omega \ll
\Delta$, the sum $\sum\limits_{\{m\}_{N-1}^n}\prod\limits_{m \in
{{\{m\}_{N-1}^n}}}f_m $, which runs over all possible $(N-1)$
single-particle levels except $|n\rangle$, does not strongly
depend on the particular level, which is eliminated from it.
Therefore in Eq.~(\ref{nBCS1}), it can be treated as a constant
and determined from the self-consistency equation $\int dx \langle
n(x,t)\rangle = 2N$, which leads to the following result
\begin{equation}
\label{nBCS} \langle n(x,t)\rangle= 2 N
\frac{\sum\limits_{n=0}^\infty f_n
\psi_n^2[x,\omega(t)]}{\sum\limits^{\infty}_{n=0}f_n}.
\end{equation}
We see that in contrast to the  case of a Fermi gas, the sum does
not have a sharp cut-off at the Fermi level. A smooth cut-off is
provided by the  factor $f_n = {v_n^2}/{u_n^2}$. E.g., if we
assume the following expressions for the variational parameters
\begin{equation}
\label{f} f_n =\frac{1 -
\frac{\varepsilon_n}{\sqrt{\varepsilon_n^2 + \Delta^2}}} {1 +
\frac{\varepsilon_n}{\sqrt{\varepsilon_n^2 + \Delta^2}}},
\end{equation}
the sum will be regularized at large indices $n \gg
\Delta/\omega$, as $ f_n \sim 1/n^2$ [we assume here that $\Delta$
is of the order of the bulk superconducting gap, $\Delta \sim
N^{1/d} \omega \exp(-1/\lambda)$, where $d$ is the
dimensionality]. Note that the only qualitative difference between
the time-dependent density profile of an expanding Fermi gas
(\ref{nFL}) and the BCS state (\ref{nBCS}) is the absence of
oscillatory features in the latter. However, we should note that
these Friedel oscillations are also absent if the expanding phase
is initially a strongly interacting but unpaired Fermi liquid.
Indeed in this case, the initial state is a Fermi sea of Landau
quasiparticles, which are not the original fermions. Therefore, if
the typical interaction energy (or temperature) is larger or of
order $\omega$, the oscillatory features are smeared out.
Therefore, the time-dependent density of a Fermi liquid state and
a paired state are qualitatively indistinguishable.

To derive the correlation function, we need to calculate the
following average $\left\langle {\rm PBCS} \right|
\hat{a}^\dagger_{k\sigma} \hat{a}_{l \sigma} \hat{a}^\dagger_{m
\sigma'} \hat{a}_{n\sigma'} \left| {\rm PBCS} \right\rangle$. We
note here that the anomalous pair correlators trivially vanish
$\left\langle {\rm PBCS} \right| \hat{a}_{l
\sigma}\hat{a}_{n\sigma'} \left| {\rm  PBCS} \right\rangle = 0$,
because the operators in the pair correlator do not conserve the
number of particles while the PBCS wave-function does so by
construction. This however does not imply that the quartic average
has no anomalous terms. This is due to the fact that the quartic
and higher-order averages  can not be decomposed into pairs using
Wick's theorem if the underlying Hamiltonian is interacting.
However, the quartic average can be calculated directly using
Eq.~\ref{PBCS}).

At this point, we introduce the following notation
\begin{equation}
\label{alpha} \alpha_{n}(x,t)= e^{in\varphi(t)}\psi_n\left[
x,\omega(t) \right],\,\,\, \mbox{ where }\,\, \tan\varphi(t)
=\omega t.
\end{equation}
This function differs from the usual oscillator wave-function only
by a time-dependant phase-factor. Using this notation, we can
write in the limit $\omega \ll \Delta$,
\begin{multline}
\label{nnBCS} \langle n(x_1)n(x_2)\rangle=\delta(x_1-x_2)\langle
n(x_1)\rangle+2\frac{Z_{N-1}}{Z_N}\left|\sum_n\alpha_{n1}\alpha_{n2}f^{1/2}_n\right|^2+
\\+\frac{Z_{N-2}Z_N}{2Z^2_{N-1}}\langle
n(x_1)\rangle\langle
n(x_2)\rangle-\frac{2Z_{N-2}}{Z_N}\left|\sum_n\alpha^{*}_{n1}\alpha_{n2}f_n\right|^2,
\end{multline}
where $Z_N = \sum\limits_{\{n\}_{N}}\prod\limits_{n
\in{\{n\}_{N}}}f_n$. The second term in Eq.~(\ref{nnBCS})
represents the ``anomalous'' component of the density, which leads
to the anomalous spatial correlations in the large-$t$ asymptotic
behavior discussed previously by Altman et al.~\cite{Altman_etal}.
Let us study the dynamics of this anomalous correlator in detail
\begin{equation}
\label{nn_an}
 \langle n(x_1,t)n(x_2,t)\rangle_{\rm anom} =
2\frac{Z_{N-1}}{Z_N}\left|\sum_n\alpha_{n}(x_1,t)
\alpha_{n}(x_2,t) f^{1/2}_n\right|^2.
\end{equation}
At $t=0$, the phase factor in the function $\alpha_n(x,0)$
(\ref{alpha}) is equal to zero $\phi(0) = 0$. If there were no
factor $\sqrt{f_n}$ in the sum in Eq.~(\ref{nn_an}), the sum would
reduce to $\sum_n \psi_n(x_1) \psi_n(x_2)   = \delta(x_1-x_2)$,
which is a same-point delta-function via the resolution of unity.
The factor $\sqrt{f_n}$, which contains information about pairing,
regularizes the sum at large indices and therefore smears out the
delta-function leading to a peak of finite width. This has clear
physical explanation: At $t=0$, the system is in the PBCS paired
state with fermions $|n,\uparrow\rangle$ and
$|n,\downarrow\rangle$ paired up. The considered limit $\omega \ll
\Delta$ implies that the size of a Cooper pair (coherence length,
$\xi_{\rm CP} \sim v_{\rm F}/\Delta \sim \sqrt{{\omega N}/{m
\Delta^2}}$) is much smaller than the size of the trapping
potential at the Fermi level $\xi \sim \sqrt{{N}/{m \omega}}$. In
our model, the coherence length  is small $\xi \gg \xi_{\rm CP}$
but finite and this leads to a finite width of the peak in the
correlation function.
\begin{figure}[htbp]
\centering
\includegraphics[width=3in]{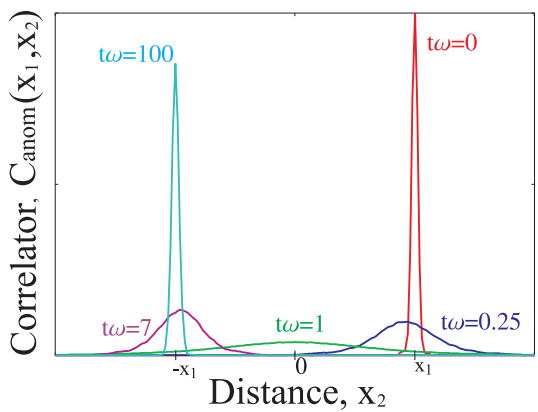}
\caption{(Color online)~The figure shows the anomalous
density-density correlator [see Eq.~(\protect\ref{nn_an})] in
rescaled units: $C_{\rm anom} (x_1,x_2;t) =
\sigma^2(t)\left\langle n[x_1 \sigma(t),t]n[x_2 \sigma(t),t]
\right\rangle$, where $\sigma(t) = \sqrt{1 + \omega^2 t^2}$. The
correlator is plotted as a function of the coordinate $x_2$ with
$x_1$ fixed. Note that at $t=0$, the correlator is peaked at
$x_2=x_1$, which reflects the fact that the size of a Cooper pair
is much smaller than the typical lengthscale determined by the
trap. At intermediate times, the correlator shows a smaller peak,
which drifts toward the point $x_2 = -x_1$. At large times, the
rescaled correlator shows a sharp peak at $x_2 = -x_1$, which
corresponds to the initial anti-correlation in momentum space. }
\label{fig1}
\end{figure}
At larger times $t > 0$, we should consider the sum of type
$\sum_n \psi_n(x_1) \psi_n(x_2) e^{-in \phi(t)}$, which  no longer
has the form of a resolution of unity. The behavior of the
anomalous correlator (\ref{nn_an}) was investigated numerically
(see Fig~1.) using the Ansatz (\ref{f}).  We note that Fig.~1
shows the evolution of the anomalous peak in terms of rescaled
variables as defined in the caption. From Fig.~1, we see that
after the expansion $0< \omega t \ll \Delta/\omega$, the initial
peak at $x_1 \sim x_2$ becomes smaller and moves toward
$x_1\sim-x_2$. At large times $t \gg \Delta/\omega^2$, the peak
re-appears exactly at $x_1 = -x_2$ in accordance with
Ref.~[\onlinecite{Altman_etal}]. In our formalism, the origin of
the peak is due to the fact that the phase factor in
Eq.~(\ref{alpha}) approaches $\phi(t\to\infty) = \pi/2$ at large
times. Therefore the sum in the anomalous correlator reduces to
$\sum_n (-1)^n \psi_n(x_1) \psi_n(x_2) \sqrt{f_n}$. Using the
symmetry properties of the Hermite polynomials, $H_n(-x) = (-1)^n
H_n(x)$, one can see that in the absence of the factor
$\sqrt{f_n}$, the sum reduces to a delta-function
$\delta(x_1+x_2)$ at the opposite points. The ``BCS factor''
$\sqrt{f_n}$ smears out the delta-function singularity; again due
to a finite size of a Cooper pair in the original condensate.


 To summarize, we have derived analytic
expressions for the dynamics of particle density and
density-density correlation function following a release of a
Fermi system from a harmonic trap. We considered two types of
initial states: A Fermi gas and a paired BCS state. Our results
for the paired state are valid only in the limit of small level
spacing $\omega \ll \Delta$. The opposite limit (in which the
Cooper pair size is larger than the typical length-scale of the
trapping potential) should be similar to the case of an
ultra-small superconducting grain (see [\onlinecite{von_Delft}]
and references therein). In the latter case, there is no true
superfluidity, but there exist a variety of interesting mesoscopic
correlation effects (see, e.g., Ref.~[\onlinecite{ML}]), which are
in principle accessible in small atomic systems.

{\em Acknowledgements ---} P.N. was supported by the JQI Graduate
Fellowship. The authors are grateful to Carlos Sa de Melo, Vito
Scarola, Eite Tiesinga, and Chuanwei Zhang for helpful
discussions.

\vspace*{-0.2cm}
\bibliography{density}

\end{document}